\documentclass[a4paper,12pt]{article}
\usepackage{color}
\usepackage[utf8]{inputenc}
\usepackage[english]{babel}
\usepackage{amssymb}
\usepackage{amsmath}
\usepackage{graphicx}
\usepackage{amsfonts}
\usepackage{latexsym}
\usepackage{epsf,epsfig,rotating}
\usepackage{cite}

\title{
\begin{flushright}
{\small INR-TH-2016-045}
\end{flushright}
Radiative corrections and instability of large Q-balls}
\author{A.\,V.\,Kovtun$\thanks{{\bf e-mail}:
andreykpq@gmail.com}$ , E.\,Ya.\,Nugaev$\thanks{{\bf e-mail}:
emin@ms2.inr.ac.ru}$ \ \vspace{10pt}\\ \emph{Institute for Nuclear Research of the Russian Academy of Science,} \\ \emph{60th October Anniversary prospect 7a, Moscow 117312, Russia} \vspace{8pt}
\\ \emph{Moscow Institute of Physics and Technology,}\\
\emph{9 Institutskiy per., Dolgoprudny,
Moscow Region, 141701,
Russia} }
\begin{document}
\maketitle
%\setlength{\oddsidemargin}{0.0\textwidth}
%\setlength{\evensidemargin}{0.0\textwidth}
%\setlength{\textwidth}{1.4\textwidth}
%\twocolumn[\maketitle
\begin{center}
\begin{abstract}
We discuss stability of Q-balls interacting with fermions in theory with small
coupling constant $g$. We argue that for configurations with large global $U(1)$-charge $Q$ the problem of classical stability becomes more subtle. For example, in model with flat direction there is maximal value of charge for stable solutions with 
$Q\sim\frac{1}{g^4}$. This result may be crucial for the self-consistent consideration of Q-ball evaporation into the fermions. We study the origin of additional instability and discuss possible ways to avoid it.
\end{abstract}
\end{center}%]
\section{Introduction}
Nontopological solitons (see \cite{Lee:1991ax} for review) are extended stationary configurations of fields which can be stable thanks to conserved global charge. Coleman discussed main properties of these lumps in \cite{Coleman:1985ki} and also referred them as Q-balls. In supersymmetric theories Q-balls may provide \cite{Kusenko:1998,Enqvist:1997si} an interesting alternative to dark matter particles
in the framework of the Affleck-Dine approach to the baryogenesis 
(see also recent astrophysical constraints in \cite{Cotner:2016dhw}).
Q-balls are also interesting for gravitation waves production
in the early universe \cite{Kusenko:2008zm,Zhou:2015yfa} and different cosmological scenarios \cite{Doddato:2012ja,Hong:2016ict}.
So, there is a motivation for theoretical studies of nontopological solitons
in the presence of other fields in different models of particle physics beyond the
Standard Model.

The stability issue of nontopological solitons is the most important problem for phenomenology. As was shown in \cite{Cohen:1986ct}, interaction with light fermions results to evaporation of Q-balls. This quantum process is parametrically suppressed for large and classically stable configurations.
In this paper we revise classical stability of Q-balls by taking into account
radiative corrections due to the fermions.
We argue that radiative corrections can provide additional instability as it was
presented in \cite{{Coleman:1973jx}} for the classical vacuum. For this purpose
the model which admits the analytical classical solution in the abscense of fermions will be studied. Remarkably, this model was the first example of Q-balls in relativistic theory \cite{Rosen0} and
can be applied for theories with flat potential. Fermions provide negative corrections to the effecctive potential for large absolute values of the scalar field. However, negative second derivative of the scalar potential is the origin of classical instability for large Q-balls \cite{Nugaev:2013poa}. This effect may be crucial in phenomenology, where charge
values of order $\sim 10^{30}$ are discussed \cite{Kasuya:2015uka}.

In the previous works restrictions on the maximal charge due to the classical fields were studied, see, for example \cite{Lee:1988ag,Multamaki:2002wk,Sakai:2011wn,Tamaki:2011zza}. The quantum stability is more subtle issue \cite{Tranberg:2013cka}. Usually evaporation of Q-ball (see e.g. \cite{Tsumagari:2009zp}) is considered as the indication of quantum instability. This is a perturbative process and the decay rate can be calculated in some approximation \cite{Cohen:1986ct} (see also \cite{Multamaki:1999an} for numerical results beyond thin-wall approximation). In addition Coleman-Weinberg mechanism discussed in our work drastically changes the stability properties for configurations with  large charge. It should be stressed that solutions exist in the absence of interaction with fermions and perturbation theory on coupling constant is applicable.
 
The paper is organized as follows. In the next section we remind general properties of Q-balls and present the model studied in more details. In Section III one-loop corrections to the scalar potential due to fermions are considered. In Sec. IV we discuss restrictions on large Q-balls and present results of numerical calculations. In Conclusions we summarized restrictions on the value of the maximal charge on the branch of stable Q-balls
and consider possible solutions to avoid them.
 
\section{Classical solutions and the model}

First of all, we briefly remind some results from the theory of nontopological solitons. 
Let us consider complex field $\phi$ with the action
\begin{equation}\label{eq:S}
S[\phi,\phi^{\ast}]= \int d^{4}x\, \left( |\partial_{\mu} \phi|^2 - V(|\phi|)\right),
\end{equation}
where $V(|\phi|)$ is the $U(1)$-invariant scalar potential.

In order to obtain the Q-ball solution we fix the charge 
\[
Q=-\int d^3 x\, i\,\left( \dot{\phi}^{\ast}\phi - {\phi}^{\ast}\dot{\phi}  \right) 
\]
and then extremize the energy 
\[
E=\int d^3 x\,\left( |\dot{\phi} |^2+| \nabla \phi |^2 +V(|\phi |) \right)
\]
with respect to this condition. 
%Explicitly it looks like
%\begin{eqnarray*}
%\min_{\phi}\left( \left. H=\int d^3 x\,\left( |\dot{\phi} |^2+| \nabla \phi |^2 +\right.%\right.\right. \\ 
%\left. \left. \left.+V(|\phi |) \right) \right|_{ Q=-\int d^3 x\, i\,\left( \dot{\phi}%^{\ast}\phi - h.c. \right) } \right)\rightarrow \\
%\rightarrow \Delta \phi +\omega^2 \phi -\frac{\partial V}{\partial \phi^{\ast}}=0.
%\end{eqnarray*}
Than, for stationary configuration $\phi(x)=e^{-i\,\omega t}\phi(\vec{x})$ one can obtain
equation of motion which we will consider later.
There are several restrictions on the frequency $\omega$ and the scalar potential:
\begin{equation}\label{eq:existcond}
\min_{\phi \neq 0} \left(\frac{V(|\phi|)}{|\phi|^2}\right)<\omega^2<m^2=\left. \frac{d^2 V}{d\phi\, d\phi^{\ast}}\right|_{\phi=0}.
\end{equation}
Derivation of these conditions are explained in \cite{Coleman:1985ki,Tsumagari:2008bv,Tsumagari:2009zp}. For the minimization of energy it is reasonable to consider spherically symmetric configurations
\begin{equation}
\phi(x)=e^{-i\,\omega t} f(r),
\end{equation}
where $f(r)$ is real nodeless function. Arguments providing that form of ansatz are discussed in \cite{Coleman:1985ki}. Equations for complex scalar field and conjugated field coincide and we could rewrite them as
\begin{equation}
f''(r)+\frac{2}{r}f'(r)+\omega^2 f(r) -\frac{1}{2}\frac{\partial V(f)}{\partial f}=0.
\end{equation}

There are several properties of solutions which can be obtained analytically. Firstly,  the useful integral condition
\begin{equation}\label{eq:dedq}
\frac{d E}{d Q}=\omega,
\end{equation}
which has generalization for gauged Q-balls \cite{Gulamov:2013cra,Gulamov:2015fya} .
We have mentioned that property because it is crucial for cross-checks, especially for numerical analysis.

Also, there is generalization of criterion \cite{Vakh:1973} of classical stability for relativistic theories \cite{Friedberg:1976me,Lee:1991ax}:
\begin{equation}\label{eq:stcond}
\frac{dQ}{d\omega}\leq 0.
\end{equation}
We choose the following potential
\begin{equation}\label{eq:flat}
V(|\phi|)=m^2 |\phi|^2 \theta\left( 1-\frac{|\phi|^2}{v^2} \right)+m^2 v^2 \theta\left( \frac{|\phi|^2}{v^2}-1 \right),
\end{equation}
which can be used for theories with flat direction. Moreover, one can consider
linear stability of solutions in this model by explicit separation of variables
\cite{Gulamov:2013ema}.
On the stable branch of solutions $E \sim Q^{3/4}$ and charge could be infinitely large for $\omega\to 0$. 

Nontrivial interaction with charged two-component left-hand spinor $\chi$
can be introduced by additional Lagrangian density 
\begin{eqnarray}\label{eq:Lpc}
\mathcal{L}=\chi^{\dagger}\bar{\sigma}^{\mu}\partial_{\mu}\chi-\frac{i\,g}{2}\left( \phi\,\chi^{\dagger}\sigma^{2} \chi^{\ast} - h.c. \right),
\end{eqnarray}
where $\bar{\sigma}^{\mu}=(1,-\vec{\sigma})$ ($\vec{\sigma}$ are usual Pauli matrices) and $g$ is dimensionless coupling constant. We suppose that both $\phi$ and $\chi$ are charged only on global U(1) group because there are serious restrictions on gauged Q-balls \cite{Lee:1988ag,Gulamov:2013cra,Gulamov:2015fya,Kusenko:1997vi}.

Interaction of the form (\ref{eq:Lpc}) results to the evaporation of Q-balls \cite{Cohen:1986ct}. But in this case there is one more quantum effect due to fermions. Radiative corrections can turn down the scalar potential. In this case for the stable branch there is a possibility to obtain solution with
\begin{equation}\label{eq:Q_0}
\frac{d Q}{d \omega} = 0
\end{equation}
for large values of charge \cite{Nugaev:2013poa}. We will look for such solutions after  
careful calculation of effective action.% Eq. (\ref{eq:Q_0}) also determines
%maximal value of charge for classical gauged Q-balls \cite{Gulamov:2013cra,Gulamov:2015fya}, where the origin of instability is
%usual Coulomb repulsion \cite{Lee:1988ag}.

During our analysis we fixed the mass $m$ of the scalar field. All dimensional variables are considered with respect to $m$ and mathematically one can put it to be equal to one.
We have emphasized above that in the theories with flat direction charge of the Q-ball might be infinitely large in the limit $\omega \rightarrow 0$. 
In this case time derivative and gradients of the scalar field are additionally suppressed compared to effective potential, which is governed only by small coupling constant $g$. Also, for potential (\ref{eq:flat}) mass of the scalar field is zero for large values of $|\phi|$ and one can take into account
only fermionic corrections. For small moduli $\phi$ we assume that
all the physics is governed by scalar field and potential (\ref{eq:flat}) has small
corrections, which will be checked in the next section.

\section{Effective potential}
Now, we turn to computation of the effective potential of the scalar field in the presence of fermions. We will use formalism of the background field, see details of this method and a lot of others in \cite{Schwartz}.

First of all, we devote a few words to the formalism of the effective action. Let us define a generating functional in the presence of classical sources $J(x)$ and $J^{\ast}(x)$.
\begin{eqnarray}\label{eq:genfun}
Z[J,J^{\ast}]=e^{i\,W[J,J^{\ast}]}=\left.\left\langle 0^{+}|0^{-}\right\rangle\right|_{\scriptscriptstyle J,J^{\ast}}= \nonumber \\
= \int \mathcal{D}\phi\, \mathcal{D}\phi^{\ast}\,\mathcal{D}\chi\, \mathcal{D}\chi^{\ast}\, e^{i\int d^{4}x \left(\mathcal{L}+\phi^{\ast}J^{\ast}+\phi J\right)}.\nonumber
\end{eqnarray}
By means of Legendre transformation we turn to the classical field $\phi$. And the effective potential in this framework is
%%%%%%%%%%%%%%%%%%%%%%%%%%%%%%%%%%%%%%%%%%%%%%%%%%%%%%%%%%%%%%%%
%When $\Phi$ is constant, we write
\[
V_{eff}(\phi,\phi^{\ast})=-\, \sum_{n}\frac{1}{(n!)^{2}} \widetilde{\Gamma}^{(1,...,2n)}(0,...,0)\, |\phi|^{2n},
\]
where 
$\widetilde{\Gamma}^{(1,...,2n)}(0,...,0)$ is 1PI Green`s function in momentum space multiplied by ${\it i}$.
% we %could rewrite (\ref{eq:epsim}) denoting 1PI function as \textquotedblleft Diagram\textquotedblright
%\begin{equation}
%V_{eff}(\phi,\phi^{\ast})=-i\,\sum_{n=1}^{+\infty}\frac{1}{(n!)^{2}}\,|\phi|^{2n}\cdot \left(
%\mathrm{Diagram} \right),
%\end{equation}
%where \textquotedblleft Diagram\textquotedblright is set of connected Feynman diagrams which cannot be disconnected %by cutting a single internal line and evaluated without propagators on the external lines.
For our purpose we take into account only one loop diagrams.
%%%%%%%%%%%%%%They are shown in Fig. (\emph{You are strongly asked to believe that they will be}).
%%%%%%%%%%%%% Think, there is no neccesity to sketch these diagrams becuse it is obvious  %%%%%%%%%%%%%%%%%%%%%%%%

We can calculate each diagram in momentum space. All of them have no external momenta. That is why we need to count a combinatorial factor which arise from topologically equivalent diagrams with crossed legs. This factor is equal to $(n!)^{2}/(2n)$. Accounting of vertices and minus sign from fermionic loop results to 
\[
-i\widetilde{\Gamma}^{(2n)}(0)=-\frac{(n!)^2\,g^{2n}}{2n} \int \frac{d^{4}p}{(2 \pi)^4} \frac{\mathrm{tr} \left( p_{\mu}\sigma^{\mu} p_{\nu}\bar{\sigma}^{\nu} \right)^{n}}{(p^{2}+i\,\varepsilon)^{2n}}.
\]

The complete expression for the one-loop correction is 
\begin{eqnarray*}
V_{(1 loop)}=i\, \sum_{n=1}^{+\infty} \int \frac{d^{4}p}{(2 \pi)^{4}} \frac{1}{n}\frac{\left(g^{2}|\phi|^{2}\right)^{n}}{\left(p^{2}+i\varepsilon \right)^{n}} +\nonumber\\
+ C|\phi|^{2}+B|\phi|^{4}.\nonumber
\end{eqnarray*}
We have already included in the expression above all counterterms which are necessary in order to cancel ultraviolet divergences. One could see that the set of integrands is the Taylor series of the logarithm and $V_{\scriptstyle (1 loop)}$ takes the form
\begin{eqnarray*}
V_{(1 loop)}=i\,\int \frac{d^{4}p}{(2 \pi)^{4}} \ln\left(1-\frac{|g\,\phi|^{2}}{p^{2}+i\varepsilon} \right)+\nonumber\\ +C|\phi|^{2}+B|\phi|^{4}.\nonumber
\end{eqnarray*}
Using the dimensional regularization we will apply the following procedure. To avoid difficulties arising from logarithm, we differentiate integrand with respect to $g^{2}$ and use the condition $\left. V_{(1 loop)}(\phi,g)\right|_{g=0}=0$.
Then one can obtain
\begin{eqnarray*}
V_{\scriptscriptstyle (1 loop)}=\frac{g^{d}|\phi|^{d}}{\left(4 \pi \right)^{\frac{d}{2}}}\, \Gamma(1-\frac{d}{2})\, \frac{2}{d}+\\
+ C|\phi|^{2}+B|\phi|^{4}=\\
=\frac{g^{d}|\phi|^{d}}{\left(4 \pi \right)^{\frac{d}{2}}}\, \Gamma(2-\frac{d}{2})
\, \frac{2}{d}\, \left(1-\frac{d}{2}\right)^{-1}+\\
+ C|\phi|^{2}+B|\phi|^{4}\, \stackrel{d=4-\varepsilon}{=} \\
\stackrel{d=4-\varepsilon}{=}
\, \frac{g^{4}|\phi|^{4}}{32 \pi ^{2}}\,\left( 
 \frac{2}{\varepsilon} - \gamma +\ln \left( 4\pi \right)\right.+\\
 \left. +\frac{3}{2} - \ln \left( g^{2}|\phi|^{2} \right)+ O(\varepsilon) \right)+\\
 + C|\phi|^{2}+B|\phi|^{4}.
\end{eqnarray*}
Renormalization conditions can be fixed at some point $M \gg v$ to be of the form
\begin{equation}\label{eq:rencond1}
\left.\frac{d^{2} V_{(1 loop)}}{d\phi \, d \phi^{\ast}}\right|_{\phi=M}=0
\end{equation}
and
\begin{equation}\label{eq:rencond2}
\left.\frac{d^{4} V_{(1 loop)}}{d\phi^{2} \, d \phi^{\ast 2}}\right|_{\phi=M}=0.
\end{equation}
Then we obtain for the effective potential
\begin{equation}\label{eq:olapp}
V_{(1 loop)}= - \frac{g^{4}|\phi|^{4}}{32 \pi^{2}}\, \left( \ln \left( \frac{|\phi|^{2}}{M^{2}}\right) -3 \right) 
-\frac{g^4\, M^{2} |\phi|^{2}}{4 \pi^{2}}.
\end{equation}
%For completeness of discussion, let us consider $V_{(1 loop)}$ a bit closer. But %before we do it, we admit that it 
Let us stress that $V_{(1 loop)}$
has two local extrema. Firstly, it has the local minimum at
\begin{equation}
\left|\phi\right|=1.60\, M.
\end{equation}
And also it has the local maximum at
\begin{equation}
\left|\phi\right|=2.59\, M.
\end{equation}
\begin{figure}
	\centering
	\includegraphics[width=0.50\textwidth]{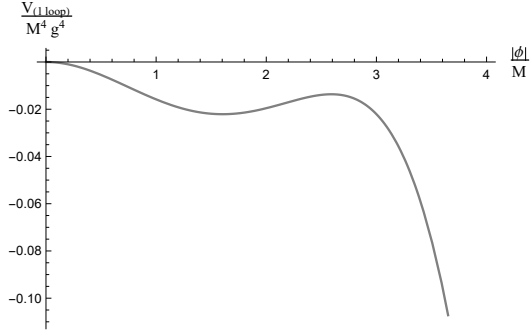}
	\caption{One-loop-correction shown in dimensionless variables.}
	\label{fig:1lc}
\end{figure}
One could introduce the new variable $\displaystyle w=\frac{\phi}{M}$ and represent  $V_{(1 loop)}$ as analytic function which scales as $g^4 M^4$ and depends on the dimensionless variable $w$
\begin{eqnarray*}
V_{(1 loop)}(\phi)= - \frac{g^{4}|\phi|^{4}}{32 \pi^{2}}\, \left( \ln \left( \frac{|\phi|^{2}}{M^{2}}\right) -3 \right)-\frac{g^4\, M^{2} |\phi|^{2}}{4 \pi^{2}}=\\
= - M^4\left(\frac{ g^{4}\left(\frac{|\phi|}{M}\right)^{4} }{32 \pi^{2}}\, \left( \ln \left( \frac{|\phi|^{2}}{M^{2}}\right) -3 \right)-\frac{g^4\,\left(\frac{|\phi|}{M^2}\right)^{2}}{4 \pi^{2}}\right)=\\
= g^{4} M^4\left(-\frac{  w^{4} }{32 \pi^{2} }\, \left( \ln \left( w^{2} \right) -3 \right)-\frac{ w^{2} }{4 \pi^{2}}\right)=\\
=g^{4}M^4\,v_{(1 loop)}(w).
\end{eqnarray*}
The dimensionless combination
\begin{equation}
v_{(1 loop)}=-\frac{  w^{4} }{32 \pi^{2} }\, \left( \ln \left( w^{2} \right) -3 \right)-\frac{\, w^{2} }{4 \pi^{2}}
\end{equation}
does not depend on parameters of the Lagrangian. In Fig. \ref{fig:1lc} one-loop correction is shown.
%Since we expect other loop corrections to be smooth with respect to %$w$, effective potential may be expressed as
%\begin{gather}\label{eq:veffexp}
%V_{\scriptstyle eff}=M^4 g^4 \left( v_{(1 loop)}+g^{n_1}v_{(2 loop)}%+g^{n_2}v_{(3 loop)}+... \right),\\
%\,1<n_1<n_2<...\nonumber
%\end{gather}
%Thus, one can consider only one-loop correction at large $\phi$, because it is possible to make $g$ small enough in order to reduce an influence of higher order corrections. Remarkably, we still can make $M$ as large as we need. 

\begin{figure}
	\centering
	\includegraphics[width=0.75\textwidth]{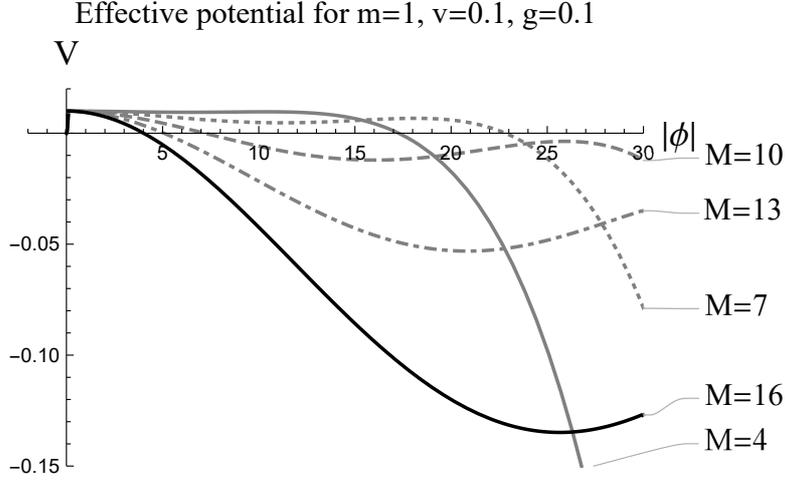}
	\caption{Effective potential with different $M$. The other free constants are g=0.1, v=0.1, m=1.0. }
	\label{fig:Vmu}
\end{figure}

In the end of the section let us look at the full effective potential which is the sum of the potential with flat direction and one-loop correction. It is of the form
\begin{eqnarray}\label{eq:effpot}
V_{eff}=m^2|\phi|^2 \theta \left(1-\frac{|\phi|^{2}}{v^{2}}\right) + m^{2} v^{2} \theta \left(\frac{|\phi|^{2}}{v^{2}}-1\right) - \nonumber \\
 - \frac{g^{4}|\phi|^{4}}{32 \pi^{2}}\, \left( \ln \left( \frac{|\phi|^{2}}{M^{2}}\right) -3 \right) 
-\frac{g^4\, M^{2} |\phi|^{2}}{4 \pi^{2}}.
\end{eqnarray}
%We sketched both potentials.
Here we should care about the classical stability of the vacuum $\phi=0$. Thus,
\begin{gather}
\left.\frac{d^2 V_{eff}}{d\phi\, d\phi^{\ast}}\right|_{\phi=0} > 0 \Rightarrow \nonumber \\
\Rightarrow m^2 > \left. \frac{d^2 V_{(1 loop)}}{d\phi\, d\phi^{\ast}}\right|_{\phi=0} \Rightarrow \nonumber \\
\Rightarrow m^2 > \frac{g^4 M^2}{4 \pi^2}. \label{eq:excond2}
\end{gather}
In Fig. \ref{fig:Vmu} the full effective potential (\ref{eq:effpot}) is presented for different values of $M$.

One could see from the Fig. \ref{fig:1lc} that one-loop correction has two slopes. The first of them starts at $|\phi|=0$ and ends at the local minimum $|\phi|=1.60\,M$. The second slope is $|\phi|>2.59\,M$. For large coupling constants one can obtain additional cusp (\ref{eq:Q_0}) on the branch of stable Q-balls when maximal value of $|\phi|$ less than $1.60\,M$ for the critical solution. The qualitative conditions for this regime are
\begin{equation}\label{eq:firstR}
\displaystyle 4.2\sqrt{ \frac{ m v}{M^2} }<g<\sqrt{\frac{2 \pi m}{M}}
\end{equation}
and will be discussed in the next section. Here we should comment that the origin of the upper bound of $g$ is the condition of the classical stability (\ref{eq:excond2}) for the vacuum $\phi=0$.

For the second regime coupling constant is small,
\begin{equation}\label{eq:secondR}
\displaystyle  4.2\sqrt{ \frac{ m v}{M^2} }>g 
\end{equation}
and the maximal value of $\phi$ for critical solution lies at the second slope.
In the next section we will derive conditions (\ref{eq:firstR}) and (\ref{eq:secondR}) in the semi-analytical approach and present numerical results.

Both regimes are illustrated at Fig. \ref{fig:Vmu}. The first regime works for the  potential with $M=16$. Curves for $M=4$ and $M=7$ correspond to the second regime.

\section{Solution for bent flat direction}
Now we turn to the discussion of solution. Following the equation of motion of the Q-ball in our model we obtain
\begin{eqnarray}
\lefteqn{f''(r)+\frac{2}{r}f'(r)+}\nonumber \\
&\displaystyle +\left(\omega^2-m^2 \theta\left( 1-\frac{f^2(r)}{v^2}  \right) +\frac{ M^2 g^4}{4 \pi^2} \right)f(r)+ \nonumber  \\
&\displaystyle +\frac{g^4 f^3(r)}{16 \pi^2}\left( \ln \left( \frac{f^2(r)}{M^2}\right)-\frac{5}{2} \right)=0.
\end{eqnarray}
We solved it numerically for two regimes (\ref{eq:firstR}) and (\ref{eq:secondR}) by the shooting method although some features can be obtained analytically.
%As most of the scientist we were trying to be on the safe side and chosen a few %particular integral conditions to make sure that solution we have obtained is the right one. First of all, it was conditions $(\ref{eq:dedq})$. And we have 
We checked that condition $(\ref{eq:dedq})$ holds for each $\omega$ for function $E(Q)$. Secondly, we have persuaded that $(\ref{eq:stcond})$ holds for stable brunch because it is crucial for our analysis. And we have made sure that $dQ/d\omega$ becomes equal to zero at the cusp point and then becomes positive. 

%As shown in Fig. \ref{fig:1lc} potential has two \emph{slopes.} \emph{Each slope could bend flat potential.} where charge constraint could occur and we continue our analysis considering two regimes.

\subsection{First regime}
%As we have mentioned it is possible to bend flat direction using the first descent of the effective potential. It is sketched in Fig. \ref{fig:veff2}. Here  we have to take care about existence condition $(\ref{eq:existcond})$. In order to provide that to be true we request

%for the existence of Q-balls.
%If it is true, Q-ball exists. 
We work in the first regime (\ref{eq:firstR}) if the amplitude of the critical solution lies within the first slope of effective potential.
\begin{figure}
	\centering
	\includegraphics[width=0.50\textwidth]{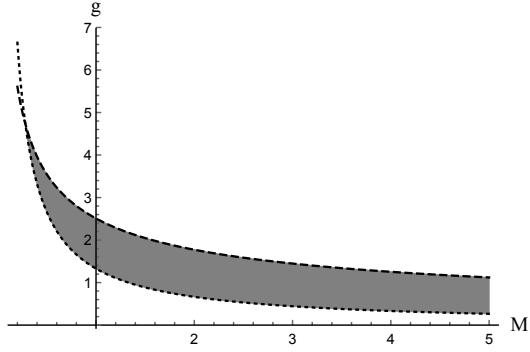}
	\caption{Domain constrained by inequalities $(\ref{eq:firstR})$ at $m=1$ and $v=0.1$.}
	\label{fig:gM}
\end{figure}
%Also we have to get another condition in order not to leave first descent of the potential. Indeed, for particular combination of free parameters solution with maximal charge may appear only in the domain of second descent. For example, $M$ dependence is sketched in Fig. \ref{fig:Vmu} shows $V_{eff}$ with different $M$, and only one of them, with $M=30$, has the solution with critical charge lying on the first descent.

%That is why it would be perfect to derive second inequality which defines %condition of leaving first descent in addition to the inequality %$(\ref{eq:excond2})$. We proceed it in approximate framework.

%When charge constraint occurs in the first descent we could approximate our  potential by simple piecewise parabolic potential of the form 

For semi-analytical analysis one can approximate effective potential by simple piecewise parabolic potential with real parameter $s$ of the form 
\begin{eqnarray}\label{eq:ppp}
V_{appr}=m^2 \left|\phi\right|^2 \theta \left(1-\frac{\left|\phi\right|^2}{v^2} \right)+\nonumber\\
+(-s^2\left|\phi\right|+(m^2+s^2)v^2 ) \theta \left(\frac{\left|\phi\right|^2}{v^2}-1 \right),
\end{eqnarray}
where $\theta(x)$ is Heaviside theta function. Solution in this potential can be obtained analytically. For large values of charge there is approximation of this analytical solution for the stable branch
\begin{equation}
\displaystyle f(r)=\left\{ \begin{array}{lr}
\displaystyle \phi_0 \frac{\sin \left(\Omega r\right)}{\Omega r}, & r<\frac{\pi}{\Omega} \\
\displaystyle 0, & r>\frac{\pi}{\Omega}
\end{array}\right. .
\end{equation} 
In this expression $\Omega^2=\omega^2+s^2$ and $\phi_0=f(0)$ is the value of the field in the centre of Q-ball. In this approximation the first term in (\ref{eq:ppp}) is negligible and one can obtain for the energy
%We choose this particular form of $\Omega$ because the largest part of energy of Q-ball is inside of radius $R=\pi/\Omega$. The last thing to do is to put $\phi(r)$ in expressions for charge and energy and find conditional minimum with respect to $\omega$. It is easy to proceed because now we have to find only local minimum of the function with a few variables.
\begin{equation}\label{eq:energy}
 E=\frac{2 \pi^2 \phi_{0}^{2}}{\Omega}+(\omega^2-s^2)\frac{2 \pi^2 \phi_{0}^{2}}{\Omega^3}+\frac{4 \pi^4 v^2 }{3 \Omega^3}(m^2+s^2),
\end{equation}
and for the charge
\begin{equation}\label{eq:charge}
Q=\frac{4 \pi^2 \omega \phi_{0}^{2}}{\Omega^3}.
\end{equation}
The conditional extremum of energy (\ref{eq:energy}) for fixed charge (\ref{eq:charge}) with respect to $\omega$  occurs for 
\begin{equation}
\phi_0=\pi v \sqrt{ \frac{m^2+s^2}{\omega^2+s^2} }.
\end{equation}
\begin{figure}
	\centering
	\includegraphics[width=0.50\textwidth]{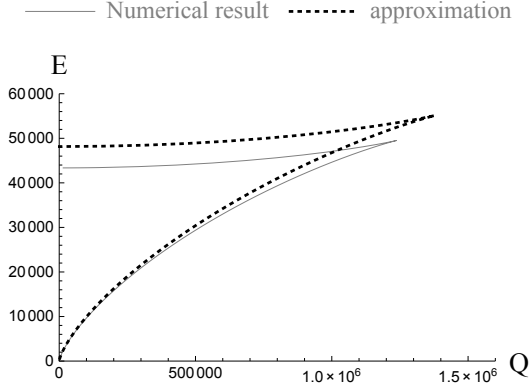}
	\caption{Parameters of the numerical solution: $g=0.1, m=1, v=0.1, M=20$. Parameters of the approximation: $m=1, v=0.1, s=0.03$.}
	\label{fig:EQ}
\end{figure}
Then the charge of configuration is 
\begin{equation}
Q=\frac{4 \pi^4 \omega v^2(m^2+s^2)}{(\omega^2+s^2)^{5/2}}.
\end{equation}
Using this expression one can obtain that condition (\ref{eq:Q_0}) occurs when
\begin{equation}
\omega=\frac{s}{2}.
\end{equation}
For this critical solution the maximal charge is
\begin{equation}
\displaystyle Q_{\max}=2 \left(\frac{4}{5}\right)^{\frac{5}{2}} \pi^4 \frac{v^2 (m^2 +s^2)}{s^4}
\end{equation}
and the field amplitude at the origin is
\begin{equation}
\phi_{0\,\max}=\frac{2 \pi v \sqrt{m^2+s^2}}{\sqrt{5} s}.
\end{equation}
Our approximate potential (\ref{eq:ppp}) is appropriate for the first regime when 
\begin{equation}
-s^2 \phi_{0 \max}^2 +(m^2+s^2) v^2 > V_{eff}(1.6 M) \nonumber
\end{equation} 
and
\begin{equation}
\displaystyle \phi_{0\,\max}< 1.6 M. \nonumber
\end{equation}

These inequalities provides the restriction 

\begin{equation}\label{eq:freg}
g>4.2\sqrt{\frac{m v}{M^2}}
\end{equation}
from the condition (\ref{eq:firstR}). The domain which responds inequalities (\ref{eq:firstR}) sketched in Fig. \ref{fig:gM}.

\begin{figure}
	\centering
	\includegraphics[width=0.50\textwidth]{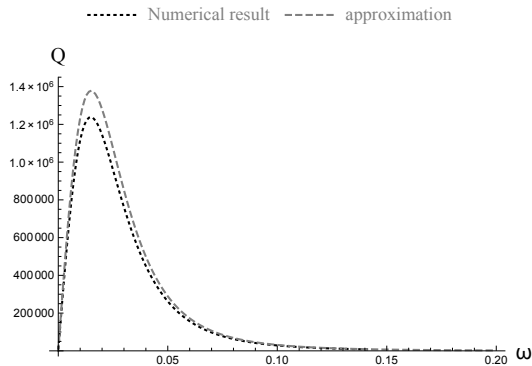}
	\caption{Parameters of the numerical solution: $g=0.1, m=1, v=0.1, M=20$. Parameters of the approximation: $m=1, v=0.1, s=0.03$.}
	\label{fig:Qom}
\end{figure}

Let us illustrate how this approximations fits numerical computations. In Fig. \ref{fig:EQ} we present how approximation holds for $E(Q)$ dependence. In Fig. \ref{fig:Qom}
two curves are plotted. Dotted curve is the numerical dependence of charge with respect to $\omega$ and dashed curve is our approximation. One could see that they almost coincide.

\subsection{Second regime}

\begin{figure}
	\centering
	\includegraphics[width=0.40\textwidth]{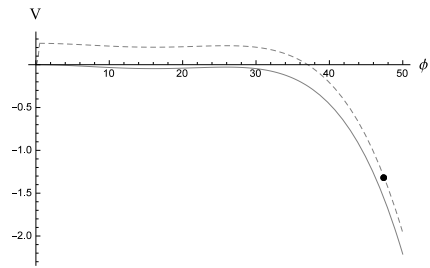}
	\caption{This plot shows the effective potential in comparison with the one-loop corrections. Black dot signs the values of the field in the center of Q-ball and value of the effective potential at the solution with maximal charge. Here $m=1, v=0.5, M=10, g=0.12$.}
	\label{fig:1veffv1l}
\end{figure}

\begin{figure}
	\centering
	\includegraphics[width=0.40\textwidth]{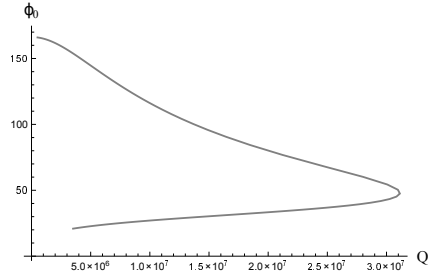}
	\caption{Plot  $\phi_{0}(Q)$ when parameters are $v=0.5,m=1,M=10,g=0.12$.}
	\label{fig:phi(Q)}
\end{figure}

In this subsection we present numerical results obtained in the second regime (\ref{eq:secondR}). In Fig. \ref{fig:1veffv1l} we signed the maximal value of the amplitude for the critical solution on the plot for the effective potential. For the same values of the parameters one can use Fig. \ref{fig:phi(Q)} to find the maximal charge for the stable branch of solutions.
%\begin{figure}
%	\centering
%	\includegraphics[width=0.50\textwidth]{EasQ}
%	\caption{Plot  $E(Q)$ when parameters are %$v=0.5,m=1,M=10,g=0.12$}
%	\label{fig:E(Q)}
%\end{figure}

\begin{figure}
	\centering
	\includegraphics[width=0.50\textwidth]{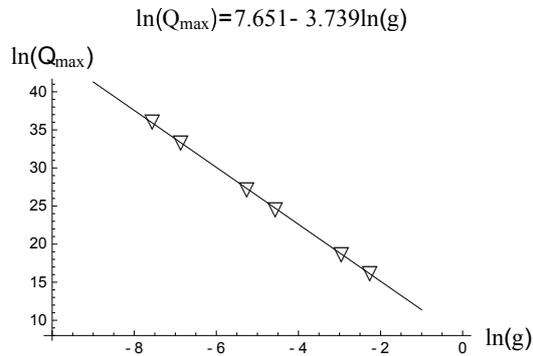}
	\caption{Plot  $\ln (Q_{max})$ versus coupling constant $v=0.1,m=1,M=1$.}
	\label{fig:lnqlng}
\end{figure}

Important result of our work is the dependence of maximal charge on the coupling constant $Q_{max}(g)$ in the second regime. We proceeded  numerical computations of this function in the limit $g \to 0$ for parameters $m=1 ,v=0.1, M=1$ and found that 
\begin{equation}\label{eq:Q(g)}
Q_{max}=\frac{2103}{g^{3.739}},
\end{equation}
as it is plotted in Fig. \ref{fig:lnqlng} in logarithmic scale.
Let us discuss analytical estimations for the $Q_{max}(g)$ of the classicaly stable solutions and compare them to numerical result (\ref{eq:Q(g)}).

% Obviously, new effect arises when flat direction starts to bend. We expect in this case that $|\phi(0)|$ for critical solution is much bigger than  renormalization scale $M$. Also one can notice that the maximum value of the field for critical solution is not too far from the point when flat direction starts to bend. That is why we are able to estimate an order of $|\phi(0)|$ for critical solution and size of Q-ball by simply finding a point $|\phi|$ where potential is equal to zero.
Here we remind the effective potential (\ref{eq:effpot}) in the limit of large $|\phi|$ and small coupling constant
\begin{equation}
V_{eff} \stackrel{ \stackrel{ g \to 0}{\,|\phi|\gg M}}{\displaystyle =} m^{2} v^{2}  - \frac{g^{4}|\phi|^{4}}{32 \pi^{2}}\, \left( \ln \left( \frac{|\phi|^{2}}{M^{2}}\right) -3 \right).
\end{equation}
We expect that in this case the maximal value of the field $|\phi(0)|$ is determined by very nonlinear part of the potential. Moreover $|\phi(0)|$ for the critical solution is of the same order as the root of the equation $V_{eff}=0$.

Assuming that logarithmic term is changing slowly and taking into account that $|\phi| \gg M$ we get that the point where potential is equal to zero is
\begin{equation}
\left.\phi\right|_{V_{eff}=0}\sim \frac{\sqrt{m v}}{g}.
\end{equation} 
%Of course we could write something like
%\[
%\left.\left(\phi^4\,\left( \ln \left( \frac{|\phi|^{2}}{M^{2}}\right) -3 %\right)<\phi^4\, 2\frac{\phi}{M}\right)\right|_{\phi\gg M}
%\]
%But it is not a big deal because in this case $\displaystyle \phi\sim %\frac{1}%{g^{\frac{4}{5}}}$ which is irrelevant with respect to our rough %analysis.
%In this framework $\displaystyle \phi|_{V_{eff}=0}$ is of the order of critical %$\phi_0$ when the charge is maximal.

In order to estimate critical charge one can use the following analysis. For sub-critical Q-balls it is possible to use flat-potential-approximation and set $s=0$ in (\ref{eq:ppp}). Then the radius of the Q-ball becomes $\displaystyle  R \simeq \frac{\pi}{\omega}$ and the maximal field amplitude $\displaystyle |\phi(0)| \simeq \frac{\pi v m}{\omega}$.
\begin{figure}
	\centering
	\includegraphics[width=0.50\textwidth]{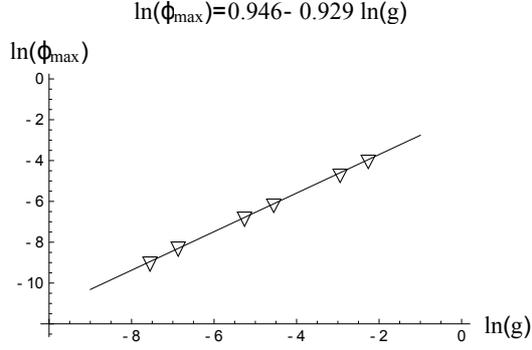}
	\caption{Plot  $\ln (\phi_{max})$ versus coupling constant $v=0.1,m=1,M=1$.}
	\label{fig:lnphilng}
\end{figure}
Then we can set 
\[
\displaystyle \phi|_{V_{eff}=0}\sim |\phi_{max}(0)| \Rightarrow \frac{\pi v m}{\omega_{crit}}\sim \frac{\sqrt{m v}}{g}.
\]
Thus, for critical solution frequency $\omega$ scales linearly with coupling constant,
\[
\omega_{crit} \sim g.
\] 
Since $\displaystyle Q \simeq R^3\,\phi_{0}^{2} \omega$
one can obtain for critical charge 
\[
\displaystyle Q_{\max} \sim \frac{1}{g^4}
\]
which is in correspondence with $(\ref{eq:Q(g)})$.
\begin{figure}[htbp]
	\centering
	\includegraphics[width=0.50\textwidth]{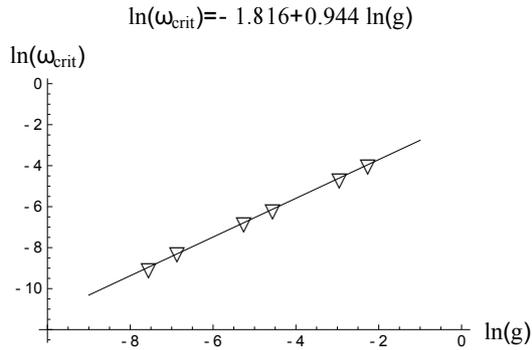}
	\caption{Plot  $\ln (\omega_{crit})$ versus coupling constant $v=0.1,m=1,M=1$.}
	\label{fig:lnOlng}
\end{figure}
On Figures \ref{fig:lnphilng},\ref{fig:lnOlng} we plotted $|\phi_{\max}(0)|$ and  $\omega_{\min}\equiv\omega_{crit}$ as functions of coupling constant $g$ in order to illustrate the validity of our approximation. It is worth to mention that the second regime is crucial for small coupling constants $g$ and the result (\ref{eq:Q(g)}) is applied for the largest Q-balls in our toy model.

\section{Conclusions}

In this work we focused on the classical stability of Q-balls in the presence of massless fermions. We assumed that scalar field has the flat direction potential which admits stable Q-balls without interactions with other fields. 
One-loop correction turned down potential for large values of moduli.
The magnitude of additional term is negative and has very nonlinear form
$\left(-\phi^4\,\ln\frac{\phi}{M}\right)$. Then we computed Q-balls solutions for different values of free parameters. Since the effective potential depends on three parameters we studied its distortions under different values of these parameters and found that charge constraint may occur in two regimes. These regimes are ruled by conditions $(\ref{eq:firstR})$ and $(\ref{eq:secondR})$.
%  It is easy to see that first regime occurs when $M \gg v$ and in this case %maximal value of the field inside Q-ball stay in the domain of the validity of %one-loop approximation. 
%In the second regime we works 
%in the limit of small coupling constants to obtain self-consistent model.
We calculate $Q_{max}(g)$ dependence when other parameters are fixed and found it to be $Q_{max} \sim \frac{1}{g^{3.739}}$. Moreover, we estimate that $\phi \sim \frac{1}{g}$ and this result is in a good agreement with numerical calculations.  %Q-ball also become too large and faces with the problem that it leaves the %domain of validity of one-loop correction. 
%According to this observation when we decrease coupling constant we may increase %renormalization scale and stay in the validity domain. 
%Also there is no problem with large logarithm since, $\phi \sim \frac{1}{g}$ and %the leading term in effective potential is proportional to $g^{0.3} \ln (g)$ %which goes to zero when $g$ goes to zero.

As a result of our work we demonstrated that interaction with fermions constrains parameters of stable Q-ball. Thus, one can not neglect this kind of interaction for large charges even if the coupling constant is very small. Our result is important for the self-consistent consideration of the Q-ball evaporation to the fermions. Restriction on the maximal charge may be strengthened or weakened because we admit interaction with other fields.
For example, gravitational coupling and electric charge also produce constraints on the charge. These effects were discussed in \cite{Multamaki:2002wk,Tamaki:2011zza,Hong:2015wga,Gulamov:2015fya}. 
As we have mentioned in the introduction, model with flat direction potential originally was considered in the framework of MSSM and now has implementations in cosmology. The breaking of supersymmetry may also rise the flat direction and this correction should be also taken into account.

\section*{Acknowledgements}
The authors are grateful to S.~V.~Demidov and M.~V.~Smolyakov for valuable discussions. The work was supported by the grant 16-12-10494 of the Russian Science Foundation.

\end{document}